# ÉTUDE DE CLASSIFICATION DES BACTÉRIOPHAGES

Dung NGUYEN[1]
Alix BOC[1]
Abdoulaye Baniré DIALLO[1,2]
Vladimir MAKARENKOV[1]

ABSTRACT – *Phages are one of the most present groups of organisms in the biosphere. Their identification continues and their taxonomies are divergent. However, due to their evolution mode and the complexity of their species ecosystem, their classification is not complete. Here, we present a new approach to the phages classification that combines the methods of horizontal gene transfer detection and ancestral sequence reconstruction.*

RÉSUMÉ – *Les bactériophages constituent l'un des groupes d'organismes les plus abondants dans la biosphère. Leur recensement est toujours en cours et les taxonomies proposées sont nombreuses et diverses. Cependant la difficulté intrinsèque, due à la diversité du mode d'évolution et à la complexité de l'écosystème des sujets, est telle qu'une classification exhaustive et convergente reste à établir. Dans cet article, nous présentons une nouvelle approche de l'étude de la classification des bactériophages. Cette approche originale combine à la fois les méthodes de détection des transferts horizontaux de gènes et de reconstruction de séquences ancestrales.*

## 1 INTRODUCTION

Les bactériophages (ou phages) sont des virus qui infectent les bactéries et les archéobactéries (ou Archaea). Leur évolution est complexe à cause des mécanismes d'évolution réticulée comprenant le transfert horizontal de gènes (THG) et la recombinaison génétique. Une représentation phylogénétique sous forme de réseau est nécessaire pour interpréter l'histoire d'évolution des bactériophages [Liu et al, 2006]. Par ailleurs, la classification de ces micro-organismes présente intrinsèquement d'autres difficultés dues, d'une part, à la non-conservation de gènes au cours de leur évolution

[1] Laboratoire de bioinformatique, UQAM, CP8888, Succursale Centre-Ville, Montréal (Québec) Canada, H3C 3P8, Nguyen.Van_Dung.2, Diallo.Abdoulaye, Alix.Boc, Makarenkov.Vladimir)@uqam.ca.

[2] McGill Centre for Bioinformatics, McGill University, 3775 University Street, Montréal (Québec) Canada, H3A 2B4.

[Rohwer et Edwards, 2002], et d'autre part, à la diversité des tailles de leurs génomes [Liu et al, 2006]. Il existe plusieurs classifications des bactériophages [e.g. Jarvis et al, 1991 ; Büchen-Osmond, 2003]. L'approche de classification, adoptée au cours des dernières décennies, est basée sur les critères de morphologie et d'homologie des ADN développés pour les phages [e.g. *L. lactis*, Jarvis et al, 1991]. La grande majorité des phages a été classée en trois principaux groupes : *936*, *c2* et *P335*. Ainsi, la plupart des études des phages tiennent compte de l'existence de ces groupes. Cependant, plusieurs travaux récents sur l'analyse comparative d'un nombre croissant de séquences génomiques et de l'émergence récurrente de nouveaux phages virulents imposent *de facto* une révision du mode de classification [e.g. Spinelli et al, 2005, Ricagno et al, 2006, Deveau, 2006]. Dans cet article, nous proposons une approche en trois phases visant à établir la classification des bactériophages : l'inférence phylogénétique, la détection de transferts horizontaux de gènes [Makarenkov et al, 2006] et la reconstruction de séquences protéiques ancestrales [Diallo et al, 2006 et Blanchette et al, 2008]. L'arbre phylogénétique des bactériophages a été reconstruit à partir d'une matrice binaire correspondant au contenu en gène de ces organismes. Les THG ont été identifiés pour déterminer le réseau phylogénétique correspondant [Makarenkov et al, 2006]. Enfin, les séquences protéiques ancestrales ont été reconstruites. Cette reconstruction ancestrale permet de mener différentes études approfondies, notamment celle concernant les origines des fonctions protéiques de ces micro-organismes.

# 2 DONNÉES SUR LES BACTÉRIOPHAGES

## 2.1 CLASSIFICATIONS EXISTANTES

Les bactériophages sont des micro-organismes très présents dans l'univers. Depuis le séquençage récent de nombreux micro-organismes, les estimations indiquent une population globale de bactériophages de l'ordre de $10^{30}$, ce qui représente la forme de vie la plus abondante sur Terre [Hendrix, 2002]. De nombreux échanges de gènes et des réarrangements de séquences à travers les recombinaisons homologues soumettent les bactériophages à une évolution réticulée [Makarenkov & Legendre, 2000; Legendre & Makarenkov, 2002; Makarenkov et al. 2004]. Le THG consiste en un échange direct de matériel génétique d'une lignée à une autre [Doolittle, 1999]. Bactéries et Archaea ont développé des mécanismes sophistiqués pour acquérir rapidement de nouveaux gènes à l'aide du THG. Le Comité International sur la Taxonomie des Virus (*International Committee on the Taxonomy of Viruses* ou *ICTV*) [Büchen-Osmond, 2003] propose une version de taxonomie de ces micro-organismes. Mais la difficulté spécifique due à la diversité du mode d'évolution réticulée et à la complexité de l'écosystème des sujets, est telle qu'une classification exhaustive et convergente n'est pas encore disponible. Par exemple, suivant les classifications établies, la majorité des phages attaquant les bactéries du lait (*Lactococcal lactis*) appartiendraient à l'un des trois principaux regroupements, notamment, *936*, *c2* et *P335* [Deveau et al, 2006]. Or, plusieurs travaux récents sur l'analyse comparative des phages semblent démontrer des incohérences dans ces regroupements [e.g. Spinelli et al, 2005, 2006; Ricagno et al, 2006; Deveau et al, 2006]. Les protéines responsables d'une même fonction dans différents organismes peuvent soit provenir d'un ancêtre commun ou d'une acquisition de gènes indépendante et spécifique à chaque lignée d'espèces [Spinelli et al, 2006]. Ainsi, les classifications des phages devraient également étudier le processus d'apparition des fonctions protéiques et l'évolution réticulée de certains gènes.

### 2.2 DONNÉES VOG

La banque de données GenBank, hébergée sur le site du *National Center for Biotehcnology Information* (NCBI) dispose d'une base de données de groupements relatifs aux protéines virales. Cette ressource, nommée *Viral COG – Clusters of Orthologous Groups* (VOG) [Bao et al, 2004], fournit des molécules standard pour la recherche génomique virale. Les données disponibles proviennent de génomes complets présents dans GenBank. Les données VOG sont des séquences de protéines regroupées de manière prédéfinie en famille selon la fonction protéique à laquelle elles sont associées. Un VOG peut comprendre des séquences de plusieurs espèces différentes. Le contenu informationnel des VOG est utilisé pour améliorer l'annotation fonctionnelle des nouvelles protéines.

L'étude phylogénétique des bactériophages présente une double difficulté en raison de la grande variabilité à la fois de la composition génétique et de la taille des génomes. La première difficulté découle de la grande divergence des séquences protéiques [Rohwer et Edwards, 2002]. La seconde difficulté est due aux tailles très disparates de génomes, qui sont d'ordre 2 de magnitude (le nombre de gènes codant en protéines varie de 8 à 381), en comparaison aux procaryotes (de ~400 à ~7 000 gènes) et aux eucaryotes (de ~4 000 à ~60 000 gènes), qui sont d'ordre 1 de magnitude [Liu et al, 2006]. Bien que la meilleure façon de normaliser les génomes de ces micro-organismes en vue d'inférer leur histoire d'évolution reste un débat ouvert [Mirkin et Koonin, 2003], la tendance actuelle est de combiner l'étude d'évolution du contenu en gènes et l'analyse des alignements de chacune des protéines qui se retrouvent dans les génomes de plusieurs phages [Liu et al, 2006]. Grâce aux VOG, les regroupements des protéines orthologues apportent des données nécessaires pour résoudre la première partie de notre problème. La méthode de normalisation de l'hétérogénéité de tailles de génomes utilisée sera présentée dans la section suivante. Dans le cadre de cette étude, 163 génomes complets de bactériophages issus de 9 familles différentes, dont une avec des annotations partielles (*unclassified*), ont été obtenus à partir de GenBank. Les séquences de ces génomes sont distribuées dans 602 regroupements VOG.

## 3 RECONSTRUCTION DE LA PHYLOGÉNIE DES BACTÉRIOPHAGES

La plate-forme d'inférence phylogénétique utilisée prend en entrée les différents groupes de VOG et les séquences protéiques associées à chaque groupe. Elle produit, en premier, un arbre phylogénétique d'espèces qui présente la première hypothèse classique sur l'évolution des bactériophages. Cette présentation en arbre ne tient pas comptes des transferts horizontaux de gènes. La plate-forme produit en second, les arbres de gènes (i.e. des protéines) individuels qui représentent l'évolution de chacun des gènes considérés dans les VOG.

### 3.1 CONSTRUCTION DE L'ARBRE PHYLOGÉNÉTIQUE D'ESPÈCES

Pour construire la classification arborescente des bactériophages, une matrice binaire de présence et d'absence de gènes (i.e. de regroupements VOG) dans chacun des 163 phages a été composée (Figure 1). Ainsi, la matrice obtenue contient 163 lignes et 602 colonnes qui représentent respectivement les phages et les VOG. Les méthodes d'inférence phylogénétique utilisant une approche de distance et une approche bayésienne ont été appliquées pour reconstruire l'arbre (voir Barthélemy et Guénoche [1988] et Makarenkov et Leclerc [1996]) pour plus de détails sur les techniques d'inférence d'arbres phylogénétiques). Des tests de robustesse ont été effectués pour mesurer le taux des regroupements d'espèces présents dans l'arbre obtenu.

### 3.1.1 Méthode de distance et méthode bayésienne

Comme requis par les méthodes de distance, une matrice de dissimilarités inter-génomiques a été initialement calculée. Plusieurs types de distances ont été récemment utilisés pour mesurer la distance entre les génomes : le coefficient de corrélation standard [Glazko et al, 2005], le coefficient de Jaccard [Glazko et al, 2005], le coefficient de Maryland Bridge [Mirkin & Koonin 2003] et la Moyenne Pondérée [Dutilh et al, 2004]. Dans cette étude, nous avons testé ces différents coefficients. Les résultats obtenus étaient très similaires, compte tenu qu'il n'y a pas d'ordre *a priori* dans les regroupements VOG. Dans notre étude, le coefficient de Jaccard a été utilisé. La méthode de Neighbor Joining (NJ) [Saitou & Nei, 1987] a permis d'inférer l'arbre phylogénétique d'espèces. Un test de bootstrap a été réalisé pour évaluer la stabilité des groupes présents dans les topologies en fonction de différents échantillons de la matrice binaire. Les différents échantillons aléatoires ont été obtenus à l'aide du programme *SeqBoot* inclus dans le paquet PHYLIP [Felsenstein, 2004]. Dans cette étude 100 échantillons ont été retenus. À la suite de l'inférence phylogénétique de chacun des échantillons, un arbre de consensus a été inféré pour chacune des approches. Le programme *Consense* inclus également dans le paquet PHYLIP a servi à générer l'arbre de consensus par la règle de majorité étendue (≥50%).

### 3.1.2 Méthode bayésienne

L'inférence bayésienne produit un arbre phylogénétique à partir de la distribution *a posteriori* des topologies d'arbres. Elle évalue l'espace de solutions au moyen des chaînes de Markov. Dans cette étude, le logiciel MrBayes [Huelsenbeck & Ronquist 2001] a été utilisé avec 2 millions de générations échantillonnées à toutes les 100 générations, 4 chaînes et 2 exécutions indépendantes, créant ainsi 20 000 arbres. Un arbre de consensus a été inféré à partir des 1000 derniers arbres les plus stables (i.e. générations stationnaires). Le programme *Consense* a servi à construire l'arbre de consensus par la règle de majorité étendue (≥50%).

## 3.2 INFÉRENCE DES ARBRES DE GÈNES

Un arbre de gène a été inféré pour chaque groupe VOG (Figure 1). Les séquences protéiques associées à un VOG donné ont été alignés en utilisant *ClustalW* [Thompson et al, 1994]. Le programme *MrBayes* a été utilisé pour inférer les arbres de gène pour les 602 alignements de séquences de VOG.

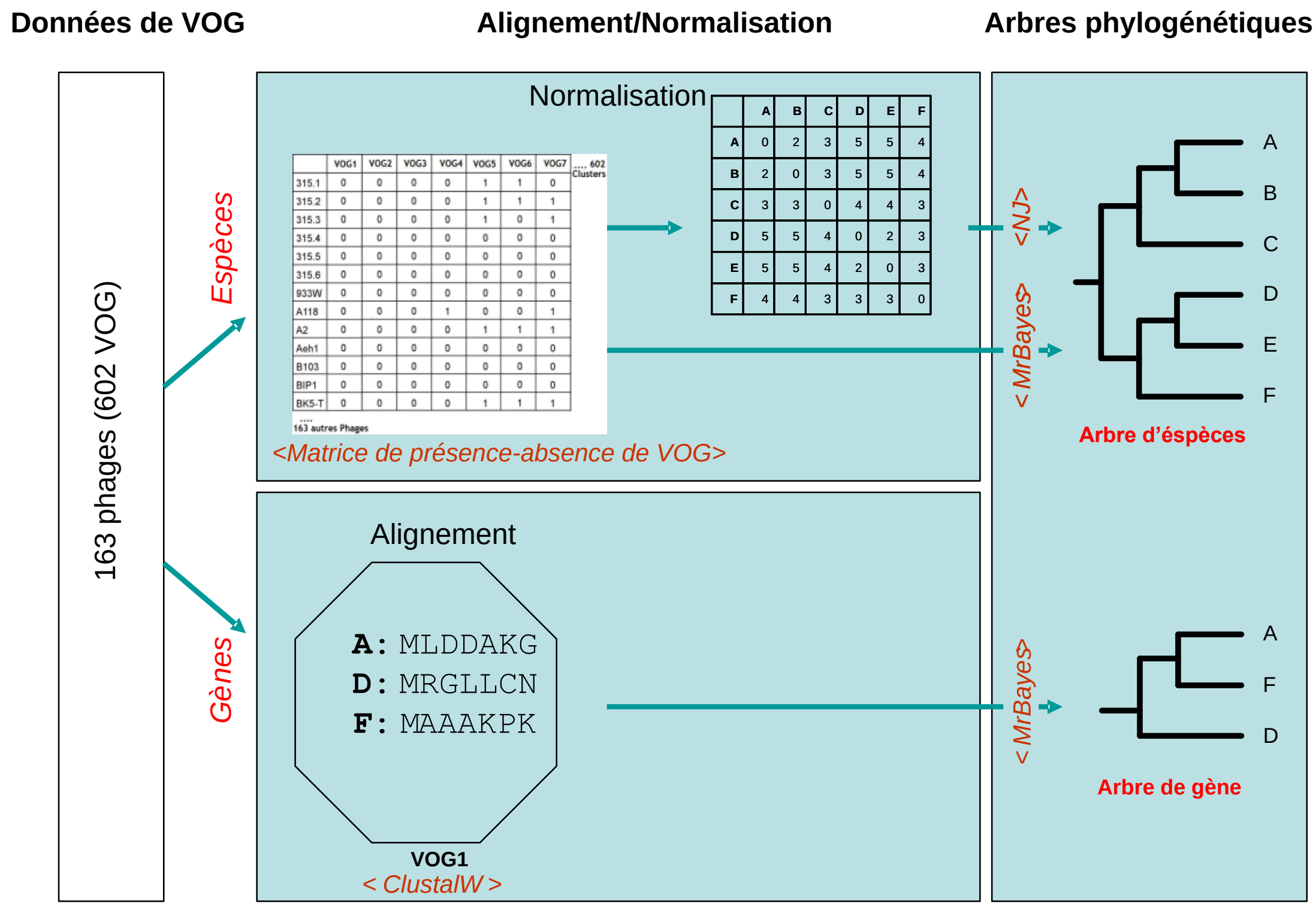

**Figure 1 : Construction des arbres phylogénétiques d'espèces et de gènes; pour les arbres d'espèces, une matrice binaire a été obtenue en fonction de la présence ou de l'absence de phage dans chaque VOG. Les arbres de gènes ont été inférés à partir des séquences protéiques alignées.**

## 4 DÉTECTION DES THG

La détection des THG a été effectuée, en utilisant le programme *HGT Detection* du package T-Rex [Makarenkov, 2001], suivant une version améliorée de l'algorithme de réconciliation topologique entre l'arbre de gènes et l'arbre d'espèces [Makarenkov et al, 2006]. *HGT Detection* (Cf. le site www.trex.uqam.ca) prend en entrée un arbre d'espèces et un arbre de gènes pour le même ensemble d'espèces. Les THG sont ainsi calculés, en indiquant en sortie l'origine et la destination pour chacun des transferts inférés. Les principales étapes de l'algorithme heuristique pour identifier des THG sont les suivantes :

*Pas préliminaire*

Inférer les arbres phylogénétiques d'espèces (i.e. arbre de contenu en gène dans notre cas) et celui de gène (l'arbre du VOG considéré dans notre cas), notés respectivement $T$ et $T'$. L'arbre $T$ est un arbre réduit de l'arbre complet construit pour 163 bactériophages et contenant seulement les phages présents dans le VOG considéré. Les deux arbres doivent être enracinés. S'il existe dans $T$ et $T'$ des sous-arbres identiques ayant au moins 2 feuilles, réduire la taille du problème en remplaçant dans $T$ et $T'$ les sous-arbres identiques par les mêmes éléments auxiliaires.

*Pas* 1 … *k*

Tester tous les THG possibles entre les paires d'arêtes dans l'arbre $T_{k\text{-}1}$ ($T_{k\text{-}1} = T$ au Pas 1) à l'exception des transferts entre les arêtes adjacentes et ceux qui violent les contraintes d'évolution (pour plus de détails voir Page et Charleston, 1998). Choisir en tant que THG optimal, le déplacement d'un sous-arbre dans $T_{k\text{-}1}$ qui minimise la valeur de la distance topologique de Robinson et Foulds [Robinson et Foulds, 1981; Makarenkov et Leclerc, 2000] entre l'arbre obtenu après le déplacement de ce sous-arbre et de son greffage sur une nouvelle arête, i.e. l'arbre $T_k$, et l'arbre de gène *T'*. Réduire ensuite la taille du problème en remplaçant des sous-arbres identiques, ayant au moins 2 feuilles, dans l'arbre transformé $T_k$ et l'arbre de gène *T'*. Dans la liste des THG retrouvés rechercher et éliminer les THG inutiles en utilisant une procédure de programmation dynamique de parcours en arrière. Un transfert inutile est celui dont l'élimination ne change pas la topologie de l'arbre $T_k$.

*Conditions d'arrêt et complexité algorithmique*

L'algorithme s'arrête quand la distance de Robinson et Foulds devient égale à 0 ou quand aucun autre déplacement de sous-arbres n'est possible suite à des contraintes biologiques. Théoriquement, une telle procédure requiert $O(kn^4)$ d'opérations pour prédire $k$ transferts dans un arbre phylogénétique à $n$ feuilles. Cependant, due à des réductions inévitables des arbres d'espèces et de gènes, la complexité pratique de cet algorithme est plutôt $O(kn^3)$.

# 5 RECONSTRUCTION DE SÉQUENCES PROTÉIQUES ANCESTRALES

La reconstruction ancestrale permet d'étudier l'évolution des espèces, la sélection adaptative et la divergence fonctionnelle [e.g. Krishnan et al, 2004]. Elle tient compte de la création de protéines et de l'évolution d'ADN en laboratoire de sorte qu'elles puissent être étudiées directement [Chang et al, 2005]. De plus, la reconstruction ancestrale de protéines peut mener aux découvertes de nouvelles fonctions biochimiques qui ont été perdues au cours de l'évolution [Jermann et al, 1995]. Les séquences de protéines renferment des informations sur leurs passés historiques [Pauling et Zuckerkandl, 1963]. Dans cette étude, nous nous sommes intéressés en particulier aux fonctions protéiques et à la recherche de séquences ancestrales des VOG. Ces séquences ancestrales faciliteraient l'analyse de similitude structurale des protéines en présentant des séquences représentatives de groupes de phages. Cette étude permet de réduire la complexité des analyses. Les séquences protéiques ancestrales et leur probabilité *a posteriori* au niveau de chaque caractère sont prédites. Ces protéines ancestrales pourraient servir comme représentants de familles de bactériophage lors de différentes analyses de génomique comparée. La reconstruction des séquences protéiques ancestrales s'effectue en deux étapes : reconstruction des ancêtres et représentation des séquences obtenues dans l'arbre d'espèces déjà construit.

## 5.1 RECONSTRUCTION DES SÉQUENCES ANCESTRALES

Au préalable, les séquences protéiques de chaque VOG ont été alignées en utilisant le programme d'alignement de séquences multiples *ClustalW* [Thompson et al, 1994]. Les arbres phylogénétiques représentant l'histoire d'évolution de chacun des VOG ont été reconstruits à l'aide du programme MrBayes [Huelsenbeck & Ronquist 2001]. L'arbre de consensus a été inféré, puis enraciné, en utilisant la technique du point médian (*midpoint*). Étant donné un alignement de séquences de régions orthologues et un arbre phylogénétique, la reconstruction de séquences ancestrales consiste à l'inférence pour chaque nœud interne de l'arbre phylogénétique, de la séquence génomique correspondante. Cette inférence s'effectue en deux étapes : la reconstruction du scénario d'insertion et de délétion (i.e. *indel*) le plus vraisemblable, et l'inférence des acides aminés à chaque position des ancêtres où la présence d'un caractère a été prédite. Ces deux étapes sont réalisées respectivement par les algorithmes de Diallo et al. [2006 et 2008] et de Felsenstein [1981] qui sont implantés dans le programme *Ancestor* disponible à l'URL suivant : <www.mcb.mcgill.ca/~banire/ancestor>.

### 5.2 REPRÉSENTATION DES SÉQUENCES ANCESTRALES

Les séquences ancestrales obtenues sont représentées sur l'arbre d'espèce (Figure 3a) en utilisant la technique de l'ancêtre commun le plus proche (ACP). Ainsi chaque VOG est associé à une séquence ancestrale située au nœud ancestral minimal commun. Il est important de noter que dans la présente étude, l'ordre des VOG n'a pas été pris en compte. Il serait intéressant dans un travail futur de trier les gènes pour déterminer l'ordre exact dans les séquences ancestrales [Bourque et Pevzner, 2002]. Cette représentation permet d'identifier au cours de l'évolution les diverses apparitions de nouvelles fonctions. Ainsi pour chaque VOG, la séquence ancestrale a été inférée. Un exemple de ces résultats présentés dans la section suivante concerne la famille des phages attaquant les bactéries du lait, *L. lactis* (sous-section 6.3).

## 6 RÉSULTATS

### 6.1 RECONSTRUCTION DE LA PHYLOGÉNIE DES BACTÉRIOPHAGES

Les arbres phylogénétiques d'espèces inférés par NJ et MrBayes étaient très similaires au niveau des regroupements d'espèces (avec de meilleurs scores de robustesse au niveau des groupes retrouvés par MrBayes). Ceci converge avec les études antérieures de comparaison d'inférence phylogénétique préconisant une meilleure précision pour des méthodes bayésiennes comparativement aux méthodes de distance [Larget et al, 2005]. Ainsi, la Figure 2 montre l'arbre phylogénétique de bactériophages avec les différentes statistiques obtenues pour la méthode bayésienne. Globalement, l'arbre phylogénétique d'espèces incorpore un grand nombre de signaux phylogénétiques : au total, 116 phages, c-à-d 71% des génomes étudiés, ont été classés dans 22 groupes avec des scores de probabilités *a posteriori* supérieur à 50%. Ces groupes robustes contiennent entre 3 et 10 phages, avec une taille moyenne de clades de 6 espèces. Plusieurs familles d'espèces, 12 sur 22 groupes, référencées par ICTV ont été retrouvées par notre analyse : *Siphoviridae* (groupes 1, 2, 6, 8, 9, 10, 13, 22), *Podoviridae* (groupes 14, 20, 21) et *Myoviridae* (groupe 4). Cependant plusieurs clades demeurent non résolus. Cela est dû à l'absence d'information convergente au niveau du contenu en gène, traduit par la présence de différentes topologies associées à ces organismes parmi les arbres générés par MrBayes. Par ailleurs, on constate que la plupart de ces clades font partie des espèces partiellement annotées (*unclassified*).

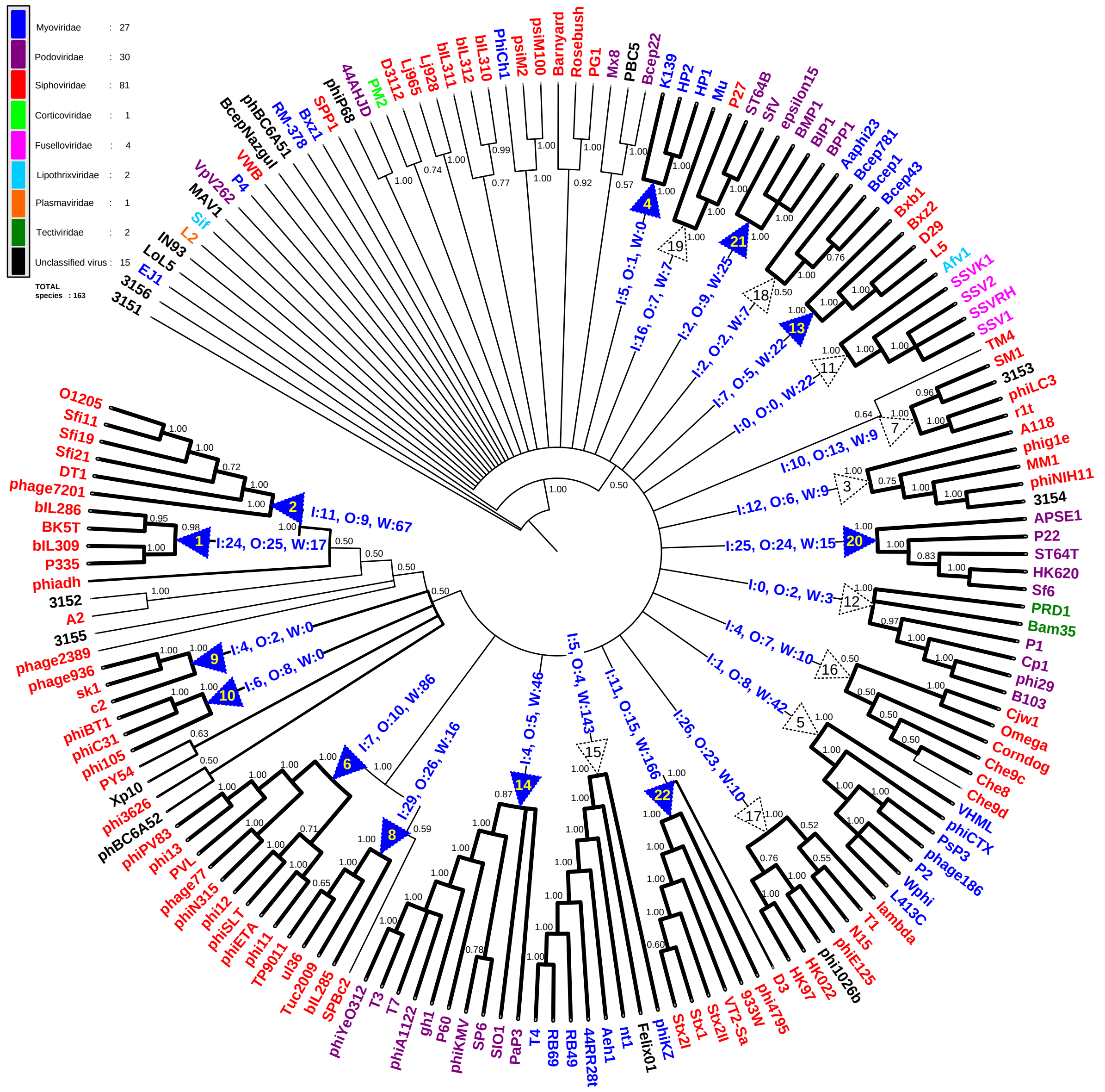

**Figure 2 : Arbre phylogénétique d'espèces inféré par MrBayes [Huelsenbeck & Ronquist 2001]. Les scores de robustesse sont indiqués pour les arêtes internes; 12 des 22 groupes identifiés (représentés par des triangles pleins) correspondent aux taxonomies de NCBI/ICTV. Pour chaque groupe, I (*In*) signifie le nombre de THG entrant dans le groupe, O (*Out*) le nombre de THG sortant du groupe et W (*Within*) le nombre de THG à l'intérieur de ce groupe. La figure a été dessinée à l'aide de l'outil de représentation d'arbres *iTol* (disponible sur : http://itol.embl.de).**

## 6.2 DÉTECTION DES TRANSFERTS HORIZONTAUX DE GENES

Au niveau des transferts, nous avons calculé les statistiques globales des THG intra (*Within*) et inter (*In/Out*) groupes. Plusieurs points sont remarquables : (a) les groupes 2, 5, 6, 12 à 16, 21 et 22 ont un nombre de transferts intra-groupes supérieur à ceux d'inter-groupes, alors que le reste des groupes a une tendance inverse, à l'exception cependant des groupes 4 et 9 qui n'ont pas de transferts intra-groupes ; (b) les groupes 1, 5, 6, 7, 10, 12, 14, 16, 21 et 22 en donnent plus qu'ils en reçoivent, et inversement pour le reste, à l'exception du groupe 11 qui ne donne ni reçoit de transferts, et du groupe 8 qui en donne autant qu'il en reçoit ; (c) les groupes qui en donnent beaucoup plus que la moyenne (informations non représentées sur la Figure 2) sont les suivants : le groupe 1 au groupe 8 (19 transferts), le groupe 17 au groupe 20 (12 transferts), le groupe 20 au groupe 17 (15 transferts) et le groupe 8 au groupe 1 (14 transferts). Les transferts entre les espèces hors groupes (i.e. les clades non résolus discutés plus haut) n'ont pas été comptabilisés dans cette étude.

### 6.3 Reconstruction des séquences protéiques ancestrales

Les résultats de la procédure de reconstruction des séquences protéiques ancestrales sont présentés sous forme d'arbres et de tableaux (Figure 3ab, vue partielle). Ainsi, nous déterminons pour chaque VOG, sa protéine ancestrale et le nœud ancestral correspondant dans l'arbre d'espèces. Ce travail permet d'identifier, à des fins de comparaison de génomes, l'ensemble des fonctions assignées à chaque nœud ancestral de l'arbre d'espèces (Figure 3b, voir les Annexes pour les résultats complets). Une prochaine étape serait la comparaison des structures des protéines ancestrales reconstruites à celles des sous-groupes correspondants. Ce travail permettrait de déterminer quels sont les domaines conservés dans les ancêtres, les fonctions existantes, les variations des séquences protéiques, les fonctions perdues par certains organismes, les fonctions acquises par les organismes de façon indépendante, etc. Ces résultats permettraient également de définir des séquences consensus pour les sous-arbres de phages, ce qui permettrait de réduire la complexité de la comparaison intergroupe de phages lors des analyses comparatives de séquences. À des fins de validation et de comparaison, il est important de mentionner ici qu'un score de prédiction de chaque séquence ancestrale (et des caractères prédits) a été également calculé en utilisant la probabilité à posteriori d'inférence de chaque caractère.

Considérons par exemple le nœud 3 (Figue 3a) qui est le nœud ancestral commun le plus proche des phages *936* (et *L. phage P2*), *c2* et *p335.* Selon la taxonomie de référence d'ICTV, les phages attaquant les bactéries de *L. lactis* sont membres de l'ordre des *Caudovirales* qui regroupe trois familles : *Myoviridae*, *Siphoviridae* et *Podoviridae*. Tous les phages attaquant les bactéries de *L. lactis* connus sont principalement membres de la famille *Siphoviridae* [Deveau et al, 2006] (commençant, en suivant l'ordre circulaire, par l'espèce *O1205* et se termine par l'espèce *SPBc2*; Figure 3a) et quelques espèces de la famille *Podoviridae* (commençant par l'espèce *phyYeO312* et se termine par l'espèce *PaP3*). Figure 3a présente un exemple d'un sous-arbre de l'arbre d'espèces complet (Figure 2) comprenant les phages attaquant les bactéries de lait. Le nœud 3 est l'ancêtre commun le plus proche des phages *P335*, *Phage936*, *c2* (et *L. phage P2*). Deux séquences protéiques ancestrales inférées pour ces organismes, leur fonction et les VOGs correspondants sont reportés dans la table (Figure 3b).

## 7 CONCLUSION

L'approche présentée ici combine à la fois les méthodes de détection de transferts horizontaux de gènes et de reconstruction de séquences ancestrales pour proposer une autre hypothèse sur la classification des bactériophages. Les résultats obtenus apportent des informations additionnelles qui visent à mieux comprendre l'histoire d'évolution de ces micro-organismes. En effet, l'issue de cette étude a permis de : (a) fournir une classification des bactériophages tenant compte de l'évolution réticulée, (b) fournir des statistiques sur les différents transferts horizontaux inter et intra-groupes, (c) générer des séquences ancestrales des phages et identifier leur origine dans l'histoire évolutive de ces organismes. Le dernier point permet aussi d'identifier des patrons communs aux groupes de séquences. Cependant, dans cette étude, nous avons occulté plusieurs problèmes dont ceux liés à l'exactitude des alignements obtenus, les scénarios de reconstructions ancestrales alternatifs ainsi que le problème lié à l'ordre des VOG dans les différents génomes. Les statistiques complètes concernant la classification des bactériophages sont disponibles à l'URL suivant : http://www.info2.uqam.ca/~makarenkov_v/Annexe_SFC2007.pdf. Dans le futur, il serait intéressant de comparer nos résultats avec ceux obtenus par Glazko et al. [2007] qui ont appliqué une méthode de classification déférente pour analyser des données de bactériophages.

a)

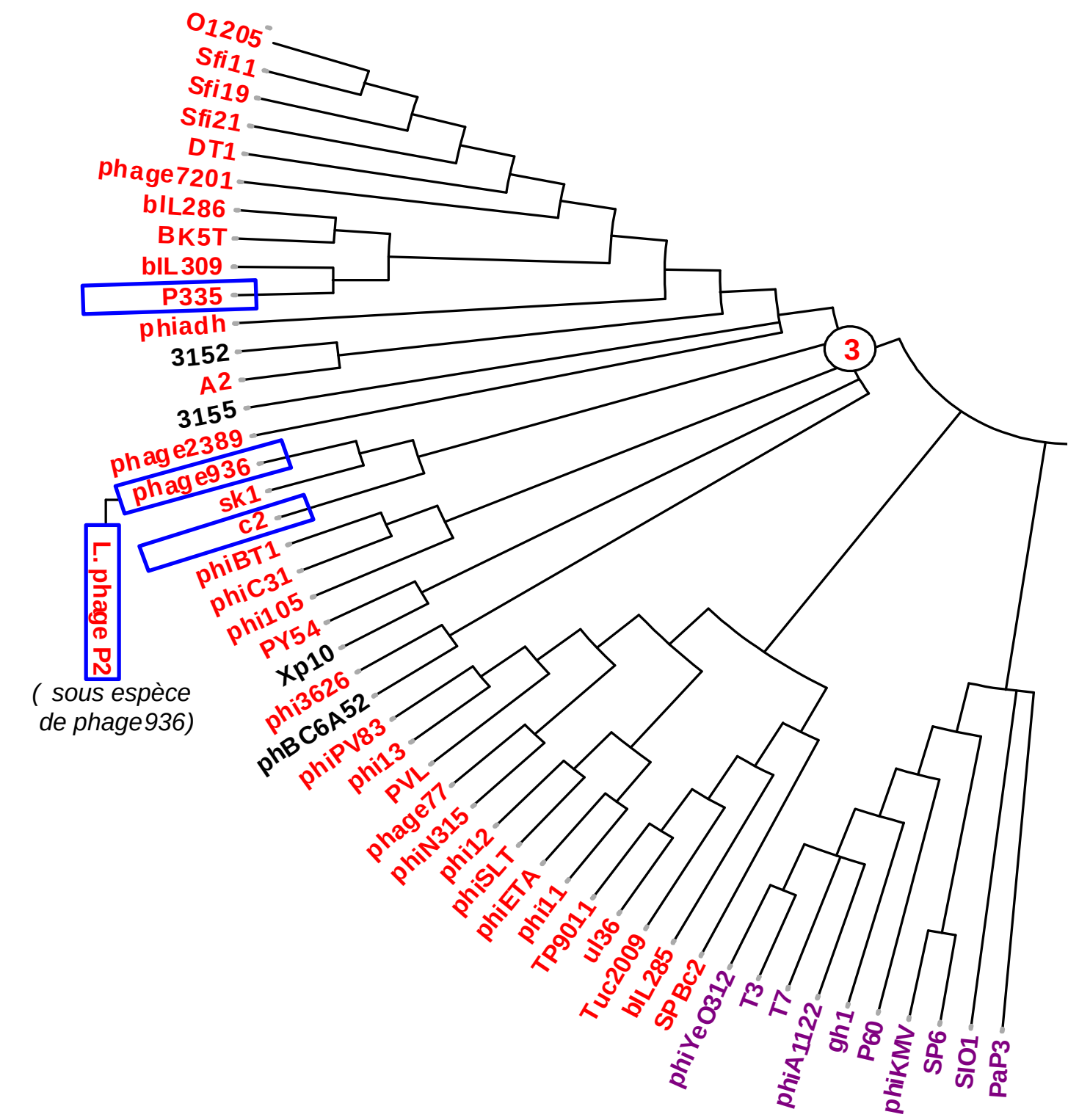

b)

| Nœud interne | Fonction | VOG | AncSeq |
|---|---|---|---|
| 3 | [T] Putative DNA binding protein | vogp0362 | EIMNRIKQLRKSRKMTRVELAEKIGVTKCTILNWEQCTSYTNPHNSQRKLADFFDVSVPYLLGZDTNZTYSNNEMALKDAIGDSIVTLVVLCLQLGYDVEECLKIAYNNIKDRQG VMN |
| 3 | [U] Uncharacterized | vogp0126 | MFNIDHSQAKDFGSIKDGTYEVIIDNANQDATKNGAEFIDIHFRIRKDFQQEFQNNNIFHRIWNDKDANKYPMAAFNNIAKAAGFPNGTKFNSLEDZLNHLLNKAFQVTVKNEKS EYKGKTYKNLNVKALAESNIPCNANPVEISEEDLPFF |

**Figure 3: (a) Sous-arbre de l'arbre d'espèces complet (Figure 2) comprenant les membres des familles *Siphoviridae* et *Podoviridae*. Le nœud 3 est l'ancêtre commun le plus proche des phages attaquant les bactéries de *L. lactis*, i.e. les organismes *P335*, *Phage936*, *c2* (et *L. phage P2*). (b) Deux séquences protéiques ancestrales inférées pour ces organismes, leur fonction et les VOGs correspondants.**

## RÉFÉRENCES